\documentclass[10pt,oneside]{article}

\usepackage{subfigure}
\usepackage{pdfpages}
\usepackage{dsfont}
\usepackage{algorithm, algpseudocode}
\usepackage{algcompatible}
\usepackage{booktabs,makecell,tabularx}
\usepackage{cite}
\usepackage{amsmath,amssymb,amsfonts}
\usepackage{graphicx}
\usepackage{textcomp}
\usepackage{xcolor}
\usepackage{bbm}
\usepackage{tabularx}
\newcommand{\algmargin}{\the\ALG@thistlm}
\makeatother
\algnewcommand{\parState}[1]{\State%
    \parbox[t]{\dimexpr\linewidth-\algmargin}{\strut\hangindent=\algorithmicindent \hangafter=1 #1\strut}}
\usepackage{tikz}
\usetikzlibrary{trees}
\newcommand{\etal}{\textit{et al}.\textrm{ }}
\tikzstyle{bag} = [align=center] 
\usepackage{multirow}
  \usepackage{cite}
\algdef{SE}[SUBALG]{Indent}{EndIndent}{}{\algorithmicend\ }%
\algtext*{Indent}
\algtext*{EndIndent}
\def\BibTeX{{\rm B\kern-.05em{\sc i\kern-.025em b}\kern-.08em
    T\kern-.1667em\lower.7ex\hbox{E}\kern-.125emX}}

\begin{document}
\title{Multi Armed Bandit based Resource Allocation in Near Memory Processing Architectures}

\author{Shubhang Pandey and
        T G Venkatesh\\Electrical Engineering Department\\Indian Institute of Technology Madras}
\maketitle
\begin{abstract}
Recent advances in 3D fabrication have allowed handling the memory bottlenecks for modern data-intensive applications by bringing the computation closer to the memory, enabling Near Memory Processing (NMP). Memory Centric Networks (MCN) are advanced memory architectures that use NMP architectures, where multiple stacks of the 3D memory units are equipped with simple processing cores, allowing numerous threads to execute concurrently. The performance of the NMP is crucially dependent upon the efficient task offloading and task-to-NMP allocation. Our work presents a multi-armed bandit (MAB) based approach in formulating an efficient resource allocation strategy for MCN. Most existing literature concentrates only on one application domain and optimizing only one metric, i.e., either execution time or power. However, our solution is more generic and can be applied to diverse application domains. In our approach, we deploy Upper Confidence Bound (UCB) policy to collect rewards and eventually use it for regret optimization. We study the following metrics- instructions per cycle, execution times, NMP core cache misses, packet latencies, and power consumption. Our study covers various applications from PARSEC and SPLASH2 benchmarks suite. The evaluation shows that the system's performance improves by $\sim11\%$ on average and an average reduction in total power consumption by $\sim12\%$.

\end{abstract}













\section{Introduction}\label{sec:introduction}
For the past few decades, the performance of a processor has observed significant improvement. In contrast, the memory performance has not been able to keep up with the processor performance improvement. The incongruency in the time taken to transfer data and the time spent to perform computations has been increasing and is referred to as the memory wall problem [1]. With the advent of modern applications which are highly data-intensive, the memory wall problem has been aggravated. Image processing applications, CNN computations, and graph processing applications need enormous amounts of data to be stored simultaneously and accessed in the minimal time possible. To address this challenge, the researchers have devised a solution to bring the computation closer to the memory [2] . 

The paradigm shift in bringing the computation closer to the memory has been possible with the evolution in fabricating 3D memory structures [3]. In this new paradigm of architecture research, two approaches have evolved significantly - Near Memory Processing (NMP)[4] and Processing in Memory (PIM) [2]. 

\begin{figure*}[t]
    \centering
    \includegraphics[width=\textwidth]{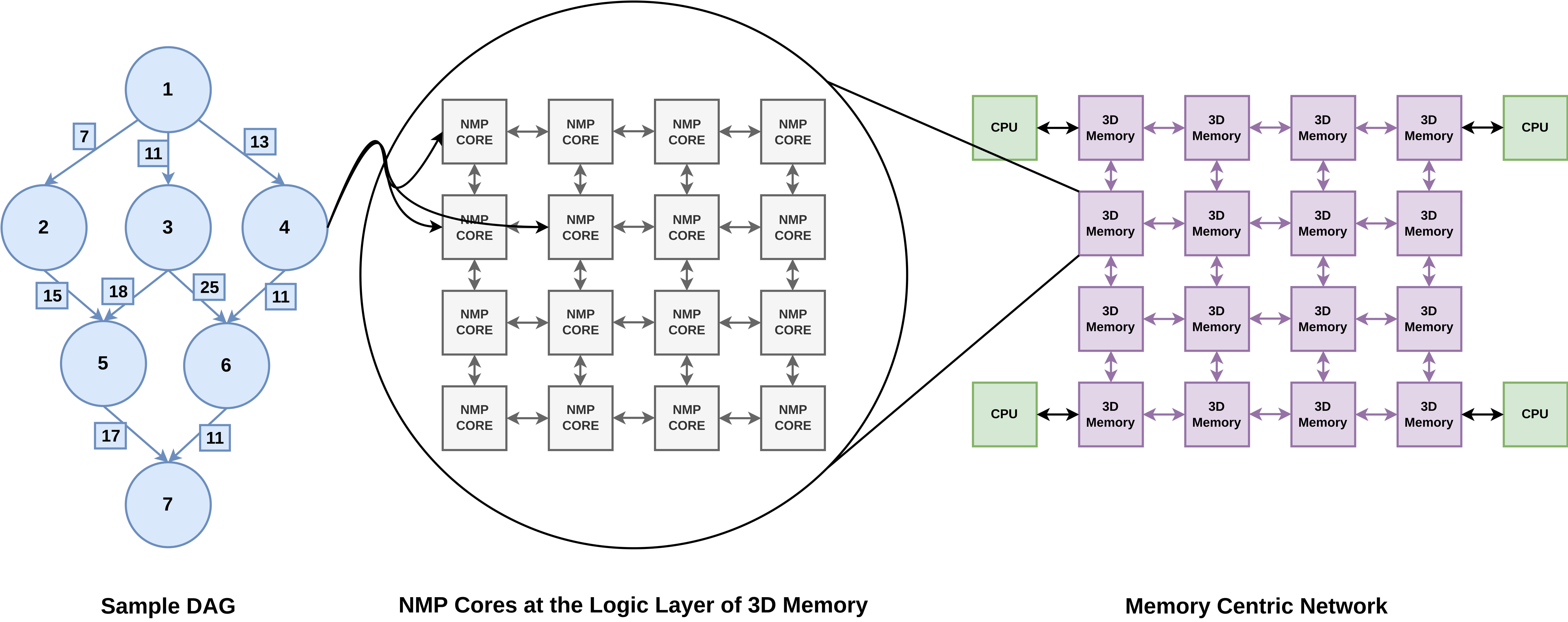}
    \caption{Memory Centric NMP Architecture implemented as Mesh Network with 16 nodes, where each node is an HMC [3]. A sample DAG with 7 nodes, where each node is a task and the corresponding edges connecting them, a detailed discussion is available in section \ref{app_exe_model}. Mapping of Application Tasks from the sample DAG to the most suitable NMP cores of the single 3D Memory of the MCN Architecture. The tasks are handled using the Tasking Model of the OpenMP [9]}
    \label{fig:paper_abstract}\vspace{-0.5cm}
\end{figure*}

The conventional CMP architectures [1] are called Processor Centric Networks (PCN). An alternate computational paradigm to the PCN is the Memory Centric Network (MCN) architecture which is more equipped to work with enormous amounts of data 
[6].  A typical example of MCN based on mesh topology is shown in fig \ref{fig:paper_abstract}. Mesh topology has several advantages, and its benefits often outweigh the costs. This architecture will remain the focus of our entire work. Each 3D-stacked memory of the MCN has an array of NMP cores. In MCN, all the communication happens only through the intermediate memory units: CPU to CPU, NMP to CPU, and CPU to NMP; all these message communications will only occur by transmitting the message across the interconnects between the memory units. 

Numerous cores available in the MCN provide an opportunity to exploit concurrent execution and maximize the performance extensively. There are two primary actions to be worked on while addressing the resource management strategies in NMP - 1) Task offloading: where a decision is taken over which sections of the application can be executed on the CPU and which can be executed on the NMP side. 2) Task allocation: which task should be executed by which NMP core is decided, i.e., mapping tasks to appropriate NMP cores. It may be lucrative to offload all the instructions to the numerous NMP cores for execution to reduce the off-chip data movement. However, this is not always beneficial for three primary reasons. 1) J. Ahn \etal [7] has established that specific applications thrive on the locality of data, and having these applications on the CPU side rather than the PIM/NMP side can improve the performance much better. 2) Performing enormous computations close to the memory may deteriorate memory performance by creating power densities. 3) The inter-task communication time becomes immensely significant when dealing with such a massive core count. Suppose the tasks are not efficiently mapped to the NMP cores. In that case, it will result in unnecessary long stall periods, under-utilized NMP cores, and overall execution delay. Given the reasons mentioned above, it becomes essential to focus on the optimal management techniques to selectively offload code to NMPs of the MCN and maintain the power profile of the NMP cores appropriately. Multi-Armed Bandit (MAB) is an evaluative feedback-based learning strategy. A major motivation in working with MAB is to have optimism in the face of uncertainty [8]. Exploration is necessary, and not always exploitation can yield optimal results. Over the years, the method has proven to be an effective way to balance resources while maximizing the benefits across various applications. 

The issue of efficient resource management in MCN  forms the motivation for our work. The paper aims to address the resource allocation problem in MCN by formulating a multi-armed bandit problem and eventually presenting performance improvement brought by our scheme in terms of execution time and power in an MCN architecture.

The remaining part of the paper is organized as follows. A brief literature survey on management methods is presented in section \ref{litsurvey}. In section \ref{model}, we briefly discuss the tasks, offloading, application model, the power consumption model. We formulate the  Multi Armed Bandit problem approach to the resource allocation strategy in section \ref{MAB}. Section \ref{sim_env} presents a discussion on the implemented methodology and the metrics used for study. Evaluation of the strategy proposed and case studies are presented in section \ref{eval}. Finally, section \ref{conclusion} presents the conclusion and the future scope of our work.

\section{Related Works} \label{litsurvey}

This section presents a brief literature survey on the development of the NMP architectures and on the various resource management methods used.

Siegl \etal [10] present a brief survey on the development of the 3D stacked memory from DRAM technology and its eventual evolution into PIM and NMP technologies. [10] highlight the various benefits in terms of latency, bandwidth and energy and emphasizes the constraints in design. Singh \etal [11] present a survey based on the potential computational units that can be used as the processing elements near memory. The processing elements observed in the survey are programmable units (simple in-order CPUs), reconfigurable units (FPGAs and CGRAs) and fixed function units (ASICs). 
Gui \etal [12] present a survey in PIM/NMP technology for graph processing applications. [12] shows that the unstructured nature in Graph processing applications benefits the most by bringing the computation closer to the memory. 

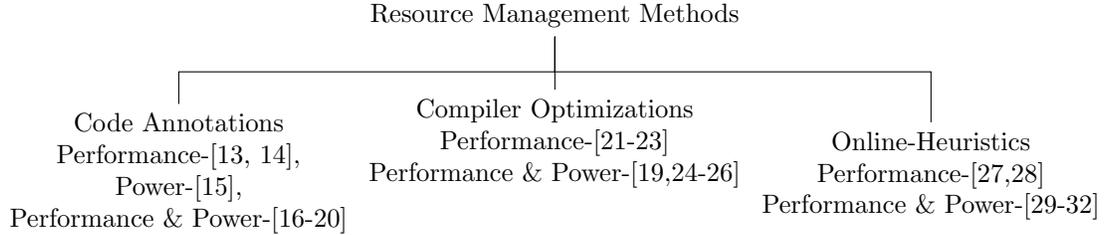
\begin{figure*}[t] 
\label{litsurtree}
\centering
\begin{tikzpicture}[level 1/.style={sibling distance=5cm}, edge from parent fork down]
 
\node[bag] {Resource Management Methods}
    child {node[bag, yshift = -.6cm] {Code Annotations\\
    Performance-[13, 14],\\
    Power-[15],\\
    Performance \& Power-[16-20]
    }}
    child {node[bag ,yshift = -0.2cm] {Compiler Optimizations\\
    Performance-[21-23]\\
    Performance \& Power-[19,24-26]}}
    child {node[bag, yshift = -.8cm] {Online-Heuristics\\
    Performance-[27,28]\\
    Performance \& Power-[29-32]\\
    }};
\end{tikzpicture}
\caption{Management Methods with Optimization Objectives}
\label{litsurtree}
\end{figure*}

Kamil Khan \etal [33] presents a detailed survey on the existing resource management methods in PIM/NMP architectures. Primary objectives of the works reported in the literature on resource management of NMPs so far include optimization for power, performance, or a combination of performance and power. As shown in figure \ref{litsurtree} resource management approach taken to achieve these objectives include-

\begin{enumerate}
    \item Code annotation ([13-20]): In this method, the programmer is expected to select appropriate portions of code to offload.
    \item Compiler optimizations ([19,21-26]): The technique identifies the appropriate sections of code to be offload during the compile time.
    \item Online-heuristics ([27,29-32]): The technique abides by the predefined set of rules to identify and offload codes during the run time.    
\end{enumerate}

Many online-heuristic approach for task offloading have been proposed in the literature. Tsai \etal [27] present thread to memory unit mapping scheduler providing improved performance with minimal overheads. Kim \etal [29] propose an online heuristic approach to selectively offload code to appropriate processing elements improving both power and performance. Lockerman \etal [30] present Livia as a hardware/software approach that places Memory Services Elements (MSEs) on the memory hierarchy. The MSEs consist of a controller and schedule the memory service tasks closest to the data. With 3\% area overhead, Livia achieves improvement in performance and reduction in dynamic energy. Choi \etal [31] focus on Image Processing application requirements and suggest a GPU-NMP design that  improves performance and reduces the hardware overhead. Pattnaik \etal [32] integrate GPU-PIM and propose methods to automatically schedule code blocks to either the PIM or GPU side and to concurrently execute the code blocks. Wang \etal [28] propose MemoNet, a memory-efficient data allocation strategy for CNN on PIM architecture. 

Some unique features of our work are as follows: 
\begin{itemize}
    \item Much effort in literature has been given to offloading the right code blocks or threads to the suitable processing unit on the NMP side. However, most of the work in task offloading is focused on specific applications such as graph processing applications, image processing, and CNN programs. However, our resource management method is more generic in nature and can be applied to a wide range of application domains.
    \item Furthermore, most of the work focuses on only one metric, which is either on performance or power. The power-performance tradeoff is often ignored. But, our work considers both power and performance, while formulating the multi-armed bandit problem.
    \item  Many works adopt a wide range of optimization and learning approaches to address the resource allocation and management problem in NMP. However, most of the above mentioned work apply a greedy strategy. Unlike others, We apply a non-greedy strategy and concentrate on long-term rewards [8]. 
\end{itemize}

\begin{enumerate}
    \item For the first time, our work establishes an evaluative feedback based learning strategy for resource allocation in NMP architectures. We formulate the resource management problem in terms of multi-armed bandit and deploy upper confidence bound policy to collect rewards. Eventually, these rewards are used for regret optimization.
    
    \item We demonstrate that our MAB formulation and regret optimization approach helps map the right tasks to the appropriate NMP cores in the MCN, thereby improving performance in terms of instructions per cycle (IPC) and execution time. 
    \item We observe NMP cache miss rates to measure the success in task allocation. We also observe the packet latency metric, thereby hinting upon the reduced contention and network traffic in the MCN architecture. 
    
    \item We also study the impact of our resource allocation strategy on reducing the power consumption of the overall system.
    
    \item Finally, We present a detailed case study on applications from two different domains. We showcase the impact of our MAB strategy on its performance over time for every group of NMP cores placed below an individual 3D stacked memory (HMC in our case) of the proposed mesh topology based MCN architecture.
\end{enumerate}

\section{Models} \label{model}
  
In this section, the first part discusses the tasks and the approach to offloading the tasks. Then, we model the execution of the application tasks and the power consumption. Later, we see that the models contribute to the various computation, communication, and power costs, which we will use to solve the resource allocation problem.

\subsection{Tasks \& Selection of Offloading Tasks} \label{l1cacheoffload}

We have used an OpenMP-based parallel programming model for the shared memory system [9]. The NMP Cores are not visible to the OS as we used the OpenMP under a shared memory model. The advantage of using OpenMP is that it is not an automatic parallel programming model and allows the programmer to control the parallelization. Traditional synchronization primitives are to be used by the programmer for the task execution. After the parallel region or region of interest completed the respective execution, task barriers were created and synchronization is done. This avoided all the side effects from race conditions, corresponding data-dependent operations, and more. The granularity of offloading the task depends on the benchmark which is being evaluated.  For example, in the ferret benchmark, the main application is to perform the image similarity search. To perform it, we have different pipeline stages. The five pipelined stages are defined as individual tasks. There is no data dependency for different images, and the pipelining can be performed easily. However, for individual operations, the respective data dependencies must be maintained [34]. An illustration of task division and allocation for the Ferret benchmark from the Parsec Benchmark is given in fig. \ref{ferret_profile} and further discussion is made available in appendix \ref{appendixA}

We follow the same offloading strategy as proposed in [7]. Ahn \etal [7] presented that the tasks within an application that is capable of exploiting locality more should be executed on the CPU side rather than the PIM side. The conventional out-of-order CPUs are more powerful than the in-order NMP cores. The cache size is smaller on the NMP side, which would generate more conflicts by default. Despite the higher performance of the CPU core, having tasks that depend less on cache support would deteriorate the system performance because of the excessive stalls due to cache misses. Such tasks which do not heavily rely on locality can be executed parallelly. It is suggested that these tasks are computed near the memory such that they take advantage of the high memory bandwidths and limit the time spent in memory accesses. Therefore, we offload the tasks by observing the L1 cache miss status as we profile the benchmarks. 

Since our focus is more on proposing an advanced task allocation strategy, having a thumb rule for the selection of tasks to be offloaded to NMP cores provides us with the opportunity to focus more on the goal of the paper. In the work done by Boromound [35], they have mentioned a way to sidestep the coherence challenge in the NMP architectures with coarse-grained granularity. 

\subsection{Application Execution Model}
\label{app_exe_model}
Following the works in [36,37], an application can be represented using a static directed acyclic graph (DAG), denoted as $\mathcal{G}=(T,E)$. A sample DAG is shown in the figure \ref{fig:paper_abstract}. $T$ is the set of $Q$ number of nodes, where each node represents a task. $E$ is the set of edges between the nodes in $T$. Each edge $e_{i,j} \in E$ represents the communication message sent from task $\tau_i$ to task $\tau_j$. The direction on $e_{i,j}$ indicates the precedence constraint that the task $\tau_i$ must be executed before task $\tau_j$ is executed. Let $M=\mu_1,\mu_2,\mu_3,\dots \mu_{\mathcal{N}}$ be the heterogeneous NMP cores ($\mathcal{N}$ denotes the number of NMP cores in the system).  Let $\theta$ be a $Q \times \mathcal{N}$ matrix, where $\vartheta_{i,k}$ is an element in $\theta$ that denotes the execution time of $\tau_i$ running on $\mu_k$. Let $\mathcal{D}$ be a matrix of dimension $Q \times Q$, where each element $d_{i,j}$ representing the size of the data in bytes to be transmitted from task $\tau_i$ to $\tau_j.$ The data transfer rates between the NMP cores are stored in matrix $\delta$ of size $\mathcal{N} \times \mathcal{N}$. The communication startup times are given in an $\mathcal{N}-$dimensional vector $\tilde{L}$, where each element $L_k$ would represent the communication startup time experienced by core $\mu_k$ in initiating the packet-based communication in the MCN architecture, when the next task $\tau_j$ on NMP core $\mu_n$ would request for the data from it predecessor task $\tau_i$ on NMP core $\mu_k$. Often if the particular memory unit is not used for a reasonable amount of time, for power-saving purposes, the memory unit is put either in the sleep, or power-down mode [3], whereby the links connecting the units too are in either of the states and eventually have to be made ready for the operation which requires some time. The communication startup time inculcates factors such as the activation time of all the necessary elements and the packet processing times (as we have considered HMCs for study) [38] [3]. The communication time of the edge $e_{i,j}$ which is for transferring data from task $\tau_i$ (scheduled on NMP core $\mu_k$) to task $\tau_j$ (scheduled on NMP core $\mu_n$) is denoted by $c_{i,j}$ and is defined by
\begin{equation} \label{task_comm}
    c_{i,j}=
    \begin{cases}
    &L_k + \frac{d_{i,j}}{\delta_{k,n}} \;\;\; \text{for} \;\;\; \mu_k \neq \mu_n\\
    &0 \;\;\; \text{for} \;\;\; \mu_k = \mu_n\\
    \end{cases}
\end{equation}

Due to the very high bandwidth offered by the TSVs in 3D memory units and the reduced data movement and high speed interconnects, it is reasonable to make two assumptions - (i) the task communication cost within an individual core is equal to 0 as shown in equation \ref{task_comm} and (ii) the communication in the network happens with negligible contention. 

 Let, $P_i$ denote the set of immediate predecessor tasks of $\tau_i$ and $S_i$ denote the set of immediate successor tasks of $\tau_i$. The set of tasks that has no predecessor task is denoted by $T_{entry}$ and the set of tasks that has no successor task is denoted by $T_{exit}$. 

After the task $\tau_j$ is scheduled on NMP core $\mu_n$, the earliest task start time (EST) is the Actual Start Time (AST) of the task $\tau_j$ on NMP core  $\mu_n$ and the earliest task finish time (EFT) is the actual finish time (AFT) of the task $\tau_j$ on NMP core  $\mu_n$.

\textbf{Definition 1:} The earliest task start time (EST) when task $\tau_j$ is scheduled for execution on core $\mu_n$ is defined as
\begin{align}
    EST(\tau_j, \mu_n)= 
    \begin{cases}
    &\max \{avail(n), \underset{\tau_i \in P_j}{max}(AFT(\tau_i)+c_{i,j})\}\\
    &0 \;\;\text{for} \;\; \tau_j \in T_{entry}\\
    \end{cases}
\end{align}

where $avail(n)$ is the earliest time at which the NMP core $\mu_n$ is available and ready for task execution. The internal maximum block in the equation returns the ready time, that is when all the data required by core $\mu_k$ is available for execution.

\textbf{Definition 2:} The earliest task finish time (EFT) when task $\tau_j$ completes execution on core $\mu_n$ as,
\begin{equation} \label{eft_cost}
    EFT(\tau_j, \mu_n)=\vartheta_{j,n} + EST(\tau_j, \mu_n)
\end{equation}

As we consider heterogeneous NMP cores, all the NMP cores should have individual resource computation rates because different cores are configured with different memory units and access operations. The focus of the model is on evaluating the resource computation cost for certain regions of the application code which can be offloaded on to the NMP cores. We define a resource computation cost rate for every heterogeneous NMP core, let $\eta=\{\eta_1,\eta_2,\dots\eta_{\mathcal{N}}\}$ be the resource computation cost rate. The resource computation cost grows constantly for each of the heterogeneous NMP core. As homogeneous communication links are considered in our work, we define a resource communication cost rate as $\zeta_{comm}$.

\textbf{Definition 3:} We define the computation and the communication resource cost $\Psi_C(\tau_j, \mu_n)$ experienced by the core $\mu_n$ while executing a task $\tau_j$ as, 

\begin{align} \label{resource_cost}
    \Psi_C(\tau_j, \mu_n) =EFT(\tau_j, \mu_n)\times \eta_n + \sum_{\tau_i \in P_j} c_{i,j} \times \zeta_{comm} 
\end{align}

here the first term contributes to the computation cost and the second term contributes to the communication cost.

\begin{table}[!htbp]
\centering
\addtolength{\tabcolsep}{-1mm}
\caption{Notations and symbols}
\label{notation}
\resizebox{\textwidth}{!}{{%

\begin{tabular}{ll}
\\

$\mathcal{C}$                                      & number of CPU cores                                                             \\
$\mathcal{N}$                                      & number of NMP cores                                                             \\
$M$                                                & set of $\mathcal{N}$ NMP cores                                                  \\
$\mu_i$                                            & NMP core $i$ in set $M$                                                         \\
$\mathcal{G}$                                      & DAG                                                                             \\
$T$                                                & set of tasks                                                                    \\
$E$                                                & set of edges                                                                    \\
$Q$                                                & number of tasks in $T$                                                          \\
$\tau_i$                                           & task $i$ from the set $T$                                                       \\
$e_{i,j}$                                          & communication edge between $\tau_i$ and $\tau_j$                                \\
$c_{i,j}$                                          & communication time for $e_{i,j}$                                                \\
$\theta$                                           & execution time matrix                                                           \\
$\vartheta{i,k}$                                   & execution time for $\tau_i$ on NMP core $k$                                     \\
$\mathcal{D}$                                      & data transfer matrix                                                            \\
$d_{i,j}$                                          & amount of data transferred from $\tau_i$ and $\tau_j$, element of $\mathcal{D}$ \\
$\delta$                                           & data transfer rate matrix                                                       \\
$\tilde{L}$                                        & communication startup time vector                                               \\
$\eta_i$                                           & resource computation cost rate for $\mu_i$                                      \\
$\zeta_{comm}$                                     & resource communication cost rate                                                \\
$h_1 ,h_2, h_3$                                    & hyperparameters associated with the costs                                       \\
EST($\tau_j, \mu_n$) & earliest task start time when  $\tau_j$ is executed on core $\mu_n$             \\
EFT($\tau_j, \mu_n$) & earliest task finish time when $\tau_j$ completes on core $\mu_n$     \\
$i_q$ & arm $i$ for task $q$ \\
$x^q_i$                                            & reward obtained from arm $i$ when task $q$ is allocated         \\
$\mathcal{J}$                                      & rewards accumulated                                                             \\
$\Delta$                                           & regret accumulated           \\ \\        
\end{tabular}%
}}

\end{table}
\subsection{Application Power Consumption Model}
Power consumption of executing a task  on a particular NMP core, denoted by $P_T$, consists of two major components - 1) Static Power $(P_{stat})$ and 2) Dynamic Power $(P_{dyn})$ [39]. $P_{stat}$ is mainly due to the leakage current and depends on the system temperature. The operational voltage and frequency contribute to $P_{dyn}$. $P_{T}$ can be shown as,
\begin{align} 
    P_{T}&=P_{dyn}+P_{stat}\\\nonumber &= {(\omega \times C_{eff} \times V_{s}^2 \times f )} \ + 
    (\alpha \times T(t) \times \beta)
\end{align}
where $\omega$ is the switching factor, $C_{eff}$ is the switching capacitance, $V_s$ and $f$ are the supply voltage and the operating frequency respectively. $\alpha$ and $\beta$ are the architecture dependent coefficients and $T$ is the chip temperature (in K) at time $t$. To formulate the power consumption model, we consider that each of the $\mathcal{N}$ NMP cores can be in one of the two states- a) Active b) Idle. $P_{stat}$ contributes to the power consumption in both the states whereas $P_{dyn}$ contributes only during the active period. 

\textbf{Definition 4:} The power consumption cost $\Psi_P(P_{T},\mu_k)$ at core $\mu_k$  be expressed as, 
\begin{align} \label{power_cost}
\Psi_P(P_{T},\mu_k)&= P_{stat} + \mathbbm{1}_k P_{dyn}\\\nonumber &\text{where} \;\; \mathbbm{1}_k=
\begin{cases}
1  \;\; \text{when core} \mu_k \\ \;\;\text{ is in operation}\\
0 \;\; \text{otherwise}
\end{cases}
\end{align}
\section{Formulation of the Multi Armed Bandit Problem} \label{MAB}

In this section, we formulate the multi-armed bandit for our resource allocation problem in MCN architecture. The costs account for the task execution time, inter-task communication, power spent in executing the task, and the resource consumption in completing the tasks. Finally, we use the regret optimization approach for task allocation in MCN using the rewards collected by the upper confidence bound (UCB) policy.

 We consider a CPU-NMP coherent design where there are $\mathcal{C}$ number of CPU cores and $\mathcal{N}$ number of NMP cores. For example with reference to fig. \ref{fig:paper_abstract}, $\mathcal{C}$ is equal to 4 and $\mathcal{N}$ is equal to 256. Both the CPU and the NMP cores share common memory space. We model the resource allocation problem as a multi-armed bandit problem, where each arm represents an individual NMP core, and pulling an arm is analogous to allocating a task to the NMP core.

 At, each play, the controller can choose to operate exactly one arm for one task. Let $x^q_i$ be the reward obtained on pulling the arm $i$ when task $q$ (similar to any task $\tau_k \in T$, defined it for the ease of expression) is being allocated, $q \in T$. At every play, the MAB strategy selects an arm $i_q \in \mathcal{N}$ and simultaneously, environment generates a reward vector $\textbf{x}^q=(x^q_1, x^q_2, \dots x^q_{\mathcal{N}})$. A play over $q$ trials consists of the sequence $(i_1, x_{i_1}^1), (i_2, x_{i_2}^2), \dots (i_{q}, x_{i_{q}}^{q})$ of the arms and the rewards selected by the algorithm and environment. Formally, in MAB, given a sequence $(i_1, x_{i_1}^1), (i_2, x_{i_2}^2), \dots (i_{q-1}, x_{i_{q-1}}^{q-1})$ of the arms pulled and rewards obtained in the first $(q-1)$ trials, the policy $\nu$ returns an arm $i_q \in \mathcal{N}$ to play in trial q. The reward of a policy $\nu$ after $k$ plays generates a sequence of reward vector $(\textbf{x}^1, \dots , \textbf{x}^k)$ given by   $\mathcal{J}_k^\nu=\sum^k_{q=1}x_{i_q}^q$.
Similarly, reward for a fixed arm $j$ of a given sequence of reward vectors $(\textbf{x}^1, \dots , \textbf{x}^k)$ is given by $\mathcal{J}_k^j=\sum^k_{q=1}x_{i_q}^q\mathbbm{1}_{i_q=j}$
 
 
We consider equations \ref{eft_cost} to represent the cost in terms of time spent in executing a task, equation \ref{resource_cost} to represent the resource consumption cost and equation \ref{power_cost} to represent the power consumption cost respectively. Mathematically, the costs are used for $x_{i_q}^q$ and can be expressed as, 
\begin{equation} \label{reward}
    x_{i_q}^q=h_1 EFT(\tau_q,\mu_i)+h_2\Psi_C(\tau_q,\mu_i)+h_3\Psi_P(P_T, \mu_k)
\end{equation}
 
 here, $h_1, h_2$ and $h_3$ are the hyperparameters. These hyperparameters are used for two main purposes- 1) to manage the costs for policy implementation and 2) to adjust the individual cost significance. It becomes absolutely essential to define the hyperparameters such that the weights on each of the rewards are applied as per requirement. Unlike the model parameters, hyperparameters can not be obtained from the data, and therefore they have to be tuned [40]. In our experiments, we performed multiple experiments for a group of benchmarks to decide upon the values $h_1, h_2$ and $h_3$.

 It may turn out better to be non-greedy and select an arm, based on the estimates for reward maximization and the possible uncertainities within those estimates. 
Here, the upper confidence bound (UCB) is the policy which we use for the MAB arm selection. 
The UCB proposes the arm selection strategy $i_q$ [8],
\begin{equation} \label{arm selection}
    i_q=\underset{j=1,\dots,\mathcal{N}}{\text{argmax}}\{\mathcal{J}^j+\sigma\sqrt{\frac{ln(q)}{n_{i_q}}} \}
\end{equation}
here, $n_{i_q}$ is the number of times the arm $i$ is played until $k$, $n_{i_q}=\sum^k_{q=1} \mathbbm{1}_{i_q=j}$. The first term is responsible for the exploitation while the second term is responsible for the exploration. The degree of exploration is guided by the hyper-parameter $\sigma$. If an arm is not pulled at all then the second term becomes very large, making it more likely to be pulled. From this we gradually become more aware of the estimate. The  MAB management method determines a  policy that maximizes $\mathcal{J}$.


Our quantity of interest is the pseudo-regret because in a stochastic framework it is more possible to achieve the optimal action in expectation than to realize the optimal action based on the rewards obtained in situ. Let $\mathcal{J}^\nu$ be the average reward accumulated when the policy $\nu$ is applied.  Let $\nu^*$ be the best arm policy, such that the reward accumulated can be denoted as $\mathcal{J^{\nu^*}}$. The ideal policy in our case would be that the tasks are executed on cores which have the minimum worst case execution time and also consume the lowest power. Then, the regret $\Delta$ can be presented as 

\begin{align} \label{regret_eq}
    \Delta&=|\mathcal{J^{\nu^*}}-\mathcal{J}^\nu|=\max_{j \in \mathcal{N}} (\mathcal{J}_{k}^j) - \mathcal{J}_{k}^\nu
\end{align}
\vspace{-0.4cm}

Algorithm \ref{PIMCoreSelect} presents a regret minimization based approach to allocating tasks to different NMP cores in an MAB based problem. The goal of the Algorithm \ref{PIMCoreSelect} is to selectively offload the tasks to the suitable NMP cores for execution such that the system benefits both on the power and performance level. The algorithm initially takes in all the information relevant to the problem in hand. The output of the algorithm is the sequence of arms pulled for every task with the minimal regret generated. The Algorithm is executed on one of the main CPU cores which is available.

With every iteration ($R$ is the total number of iterations), the rewards are collected based on the NMP cores or arms which execute the task. As mentioned, we rely on the Upper Confidence Bound policy for exploring all
the available cores and computing the rewards based on the costs as shown in equation \ref{reward}.
With every MAB run, regrets are computed as well. Regret optimization (or minimization) is our objective through out the MAB based resource allocation problem, so whenever the current regret value is less than the previously observed regret the sequence of arm selections is updated and this information is held until the next obtained regret is better than the current. Finally, over multiple iterations, we obtain a sequence of arms that generate the most optimal regret. The final selected arms sequence is nothing but the sequence of NMP cores on which the tasks are to be executed.

The affinity of the data to the execution core is essential in improving the performance of the system overall. If a particular task is executed closer to the needed data, an improvement will be clearly visible in the execution time of the benchmark. This is because, in the NMP system discussed in the paper the TSVs, connecting the bottom layer of the stack with memory provide high bandwidth compared to the on-chip links. Again, the data has to experience very less buffer time while communicating from one core to the other. Since we have considered the execution time as a reward metric in our model, the data affinity is inherently taken care. However, we experienced some exceptions to the above-mentioned scenario, where offloading the task to the next NMP core also improved the system performance, as the load was balanced across all the cores. The scenario was also well identified by the MAB-based resource allocation strategy, as it continuously strived to optimize the cumulative negative reward.   
 

With our approach, the decision to find the suitable cores for each task is taken care by the MAB algorithm. The algorithm assigns negative reward to the execution times, which means higher the execution time, higher is the cumulative reward. The fundamental objective of the algorithm is to optimize the cumulative negative reward below the regret threshold value. Thus our MAB algorithm automatically favours those task allocation policies wherein tasks are in proximity to the data they manipulate.

\begin{figure}[t]
     \centering\includegraphics[width=\textwidth]{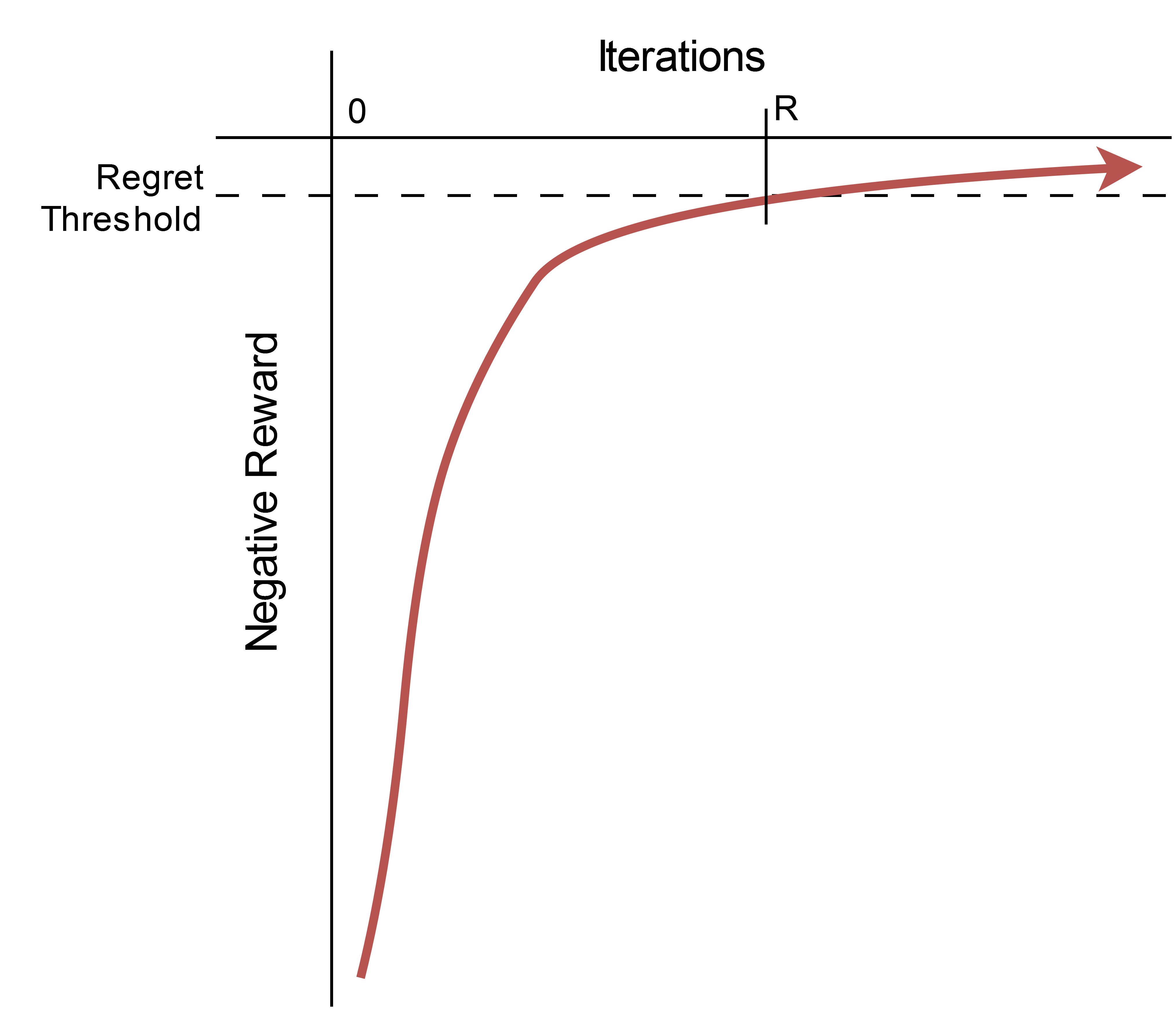}
     \caption{Example of reducing negative reward and deciding on the iteration value R}
     \label{rew_1_regretopto}
 \end{figure}

The reduction in the cumulative negative reward over the multiple iterations can be seen in the illustrative diagram shown in fig \ref{rew_1_regretopto}.


\begin{algorithm}[t]
\caption{NMP Core Selection based on regret optimization}
\label{PIMCoreSelect}
\begin{algorithmic}
\STATE \textbf{Input:} $\mathcal{G,D},\theta, \mathcal{N}, \sigma, h_1, h_2, h_3, \eta, \zeta_{comm},R$ 

\STATE \textbf{Output:} $A_t$ , $\Delta$
\STATE

\FOR{$itr=1,2, \dots R$}
\STATE Estimate  $\mathcal{J}^{\nu^*}$ // Accumulated reward under best conditions 
\STATE Apply UCB arm selection strategy $I_t$ using equation  \ref{arm selection} 
\STATE Compute the task execution cost using equation \ref{eft_cost}
\STATE Compute resource consumption cost using equation \ref{resource_cost}
\STATE Compute the power consumption
cost using equation \ref{power_cost}
\STATE Gather the reward for each arm using
equation \ref{reward} 
\STATE Update $\Delta$ based on equation \ref{regret_eq}
\STATE Append the sequence of arms $A_t$ for task execution

\IF{current $\Delta$ $\leq$ previous $\Delta$}

\STATE hold the value of current $\Delta$
\STATE Update the array $A_t$ with the sequence of arms

\ENDIF
 
\ENDFOR
\STATE \textbf{return} $A_t$, $\Delta$
\end{algorithmic}
\end{algorithm}

\section{Simulation Environment} \label{sim_env}

\begin{figure}[t]
    \centering
    \includegraphics[width=\textwidth,height=0.3\textheight]{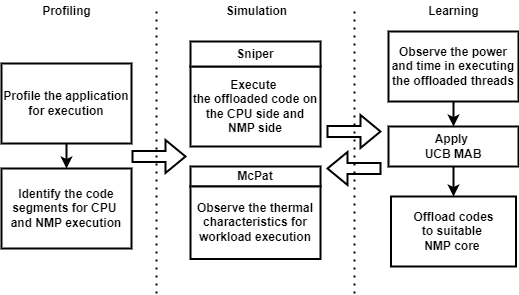}
    \caption{Implemented Methodology}
    \label{implementation}\vspace{-0.5cm}
\end{figure}

Our simulation environment consists of Multicore Out-of-Order x86 CPU with private L1 instruction and data caches, partially shared L2 cache, and shared L3 cache. We consider Hybrid Memory Cube (HMC) [3] as an example of the 3D stacked memory design; however, any other 3D stacked memory unit such as the HBM can be considered. NMP cores are simple in-order x86 cores placed below every vault in the HMC unit with L1 and L2 caches as private. For more details on the architecture configured for the simulated environment please refer the tables- CPU configuration table \ref{CPUSpec}, NMP configuration table \ref{NMPSpec}, and 3D memory specifications table \ref{hmc_spec}. We build our CPU-NMP environment on Sniper [41]. McPat [42] is tuned for the power evaluation of the CPU and NMP architectures.

\begin{figure}[t]
    \centering
    \includegraphics[width=\textwidth]{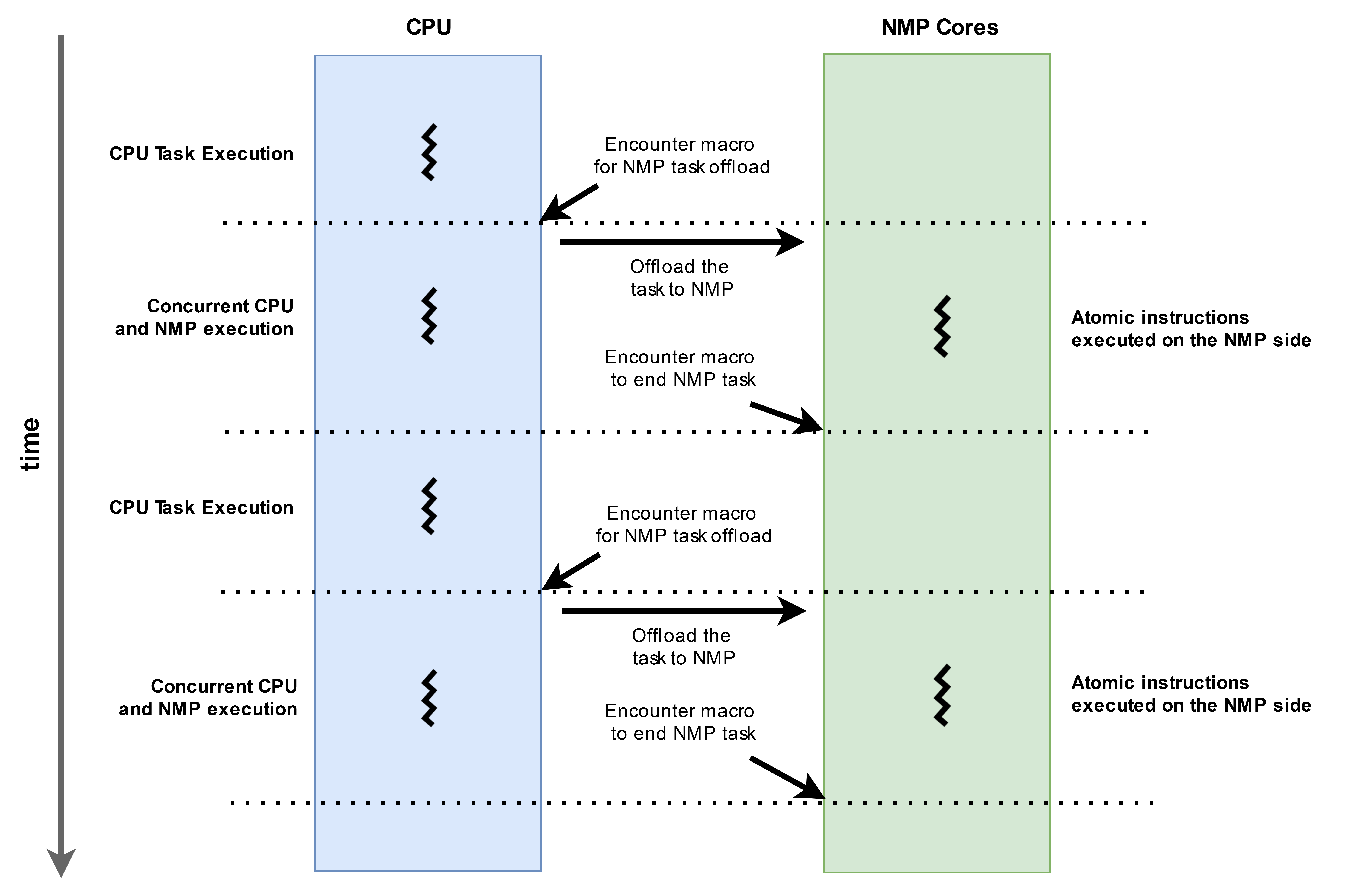}
    \caption{Task Offloading Strategy under OpenMP Programming Model}
    \label{task_offload_strat_rev_1}
\end{figure}

In fig. \ref{implementation}, a high level implementation strategy is presented. We first profile the benchmarks from the PARSEC Benchmark Suite 3.0 [43] and SPLASH2 [44] on the configuration environment on Sniper 7.3 [41] and observe the power characteristics on McPat [42] which is integrated with Sniper. We rely on the api present in Sniper to mark the regions of interests. We even validate our profiling method with the work done by Wei \etal using PIMProf [45] for the same set of benchmarks. Once the profiling is done, we use the simulation data to identify the suitable portions of code to offload. We identify the NMP cores where the defined tasks can be executed. Identifying the suitable tasks to be offloaded on the NMP for execution is a major challenge. However, addressing the issue is out of the scope of the paper. Hence, we refer to the work by Ahn \etal [7], where the author allocates the tasks which thrive less on the locality of the data to the NMP core for execution. On a similar note we identify the tasks which do not rely heavily on larger caches placed close to the CPU. Such tasks are candidates for offloading to NMP cores. The diagram shown in fig. \ref{task_offload_strat_rev_1} presents the task offloading strategy used in the paper. When the compiler encounters a hard macro under the OpenMP Programming Model, it offloads all the following tasks for execution on to the NMP cores.

The following are the metrics which we have focused upon - \textit{1) Instructions per cycle (IPC)}: It is the average number of instructions executed per clock cycle. \textit{2) Execution Time}: From here on we will call the NMP task offloaded parts as the region of interest or ROI. Execution time is the total time required to execute the ROI. \textit{3) NMP L1 Cache Miss Rates}: The miss rate is given by the ratio of number of access misses to the total number of accesses made, i.e. $\text{miss rate} = \frac{\text{number of accesses missed}}{\text{total number of accesses}}$. \textit{4) Average Packet Latency}: It is the metric to indicate the overall contention in the network. Average Packet Latency is defined as the ratio of the total packet delay over the mesh and the total number of packets. and \textit{5) Percentage Power Reduced}: Potentially all the system are evaluated on the success of its behaviour in terms of power.

\vspace{-0.3cm}
\begin{table}[!htbp]
    \centering
    \caption{CPU Configuration}
    \begin{tabular}{|c|c|} \hline
         CPU & 16 cores, Out-of-Order, 3.2GHz\\ \hline
         L1I & Private, 32KB, 4-way assoc \\ \hline
         L1D & Private, 32KB, 8-way assoc \\ \hline
         L2 & 4 core Shared, 256KB, 8-way assoc \\ \hline
         L3 & Shared, 16MB, 16-way assoc \\ \hline
         TLB & I-TLB and D-TLB: 256 entries each \\ \hline
    \end{tabular}
    
    \label{CPUSpec}
\end{table}
\begin{table}[!htbp]
    \centering
    \caption{NMP Configuration}
    \begin{tabular}{|c|c|} \hline
         NMP-CPU & 1 core, In-Order, 2GHz\\ \hline
         L1I & 16KB, 4-way assoc \\ \hline
         L1D & 16KB, 4-way assoc \\ \hline
         L2 & 128KB, 8-way assoc \\ \hline
         L3 & 4MB, 16-way assoc \\ \hline
        TLB & I-TLB and D-TLB: 128 entries each \\ \hline
        Topology & 16 node Mesh \\ \hline
    \end{tabular}
    
    \label{NMPSpec}
    \vspace{-0.5cm}
\end{table}
\begin{table}[!htbp]
    \centering
    \caption{3D Memory Specification}
    \begin{tabular}{|c|c|} \hline
    Number of Vaults \& Banks & 16 \& 16 \\\hline
    Memory Capacity & 8GB \\ \hline 
    Number of DRAM Layers & 8 layers\\ \hline
    Link Width & 4 (Quarter) \\ \hline
    Link Lane Speed &  12.5Gb/s \\ \hline
    Retry Buffer Size & 32 flits \\ \hline
    Page Policy & Closed Page Policy \\ \hline
    Vault controller buffer size & 32 flits \\ \hline
    Crossbar switch buffer size & 64 flits\\ \hline
    Maximum Aggregate Link Bandwidth & 480GB/s \\ \hline
    Maximum DRAM data bandwidth & 320GB/s \\ \hline
    Maximum Vault data bandwidth & 10GB/s \\ \hline
    \end{tabular}
    \label{hmc_spec}
\end{table}
\section{Evaluation} \label{eval}
In this section, we examine the power and performance improvement exhibited by the MAB resource allocation strategy, the metrics for this study have been discussed in section \ref{sim_env}. We also present two case studies on applications from Financial Analysis and Engineering, where the featured programs are diverse from one another.

\subsection{Evaluation of MAB strategy}
\begin{figure}
\centering
\includegraphics[width=\textwidth]{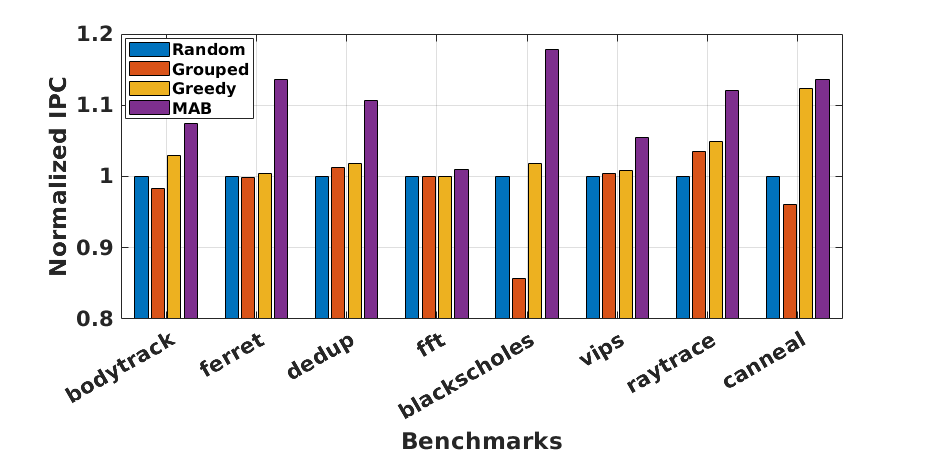}
    \caption{Performance Comparison in terms of Normalized IPC for Benchmarks from PARSEC and SPLASH2 Benchmarks Suite}
    \label{MAB_IPC}
\end{figure}

The random and grouped allocation strategies are comparison metrics introduced by us for the study and evaluation of the proposed resource allocation strategy. These metrics are not considered by others in the literature. A baseline diagram of the MCN architecture under study is given in figure \ref{fig:paper_abstract}, the diagram is needed to understand the architecture and resource allocation strategies.

Under the random allocation strategy, the offloaded NMP tasks can be mapped onto any available NMP core. The random strategy does not take into consideration of any power or performance benefit but rather allocates the tasks based on the sheer availability of the NMP core. The grouped strategy on the other hand takes into consideration the data affinity and allocates the tasks to the cores which are nearest to the data. For some benchmarks, the grouped strategies may prove to be immensely useful where data affinity is a critical factor. However, a major drawback to the grouped strategy is the creation of power densities, which directly affect the memory performance, i.e. reducing the application execution time. The greedy allocation strategy is very similar to the methods used in the literature. In the work done by Hameeza Ahmed \etal in PRIMO [23] and Ramyad Hadidi \etal in CAIRO [21], the compiler selects the best suitable tasks for PIM execution and improves the performance. However, the approach is only applicable to specific kernels and applications. Jorge Veiga \etal [22] work on improving the efficiency of Processing-in-memory  focused only on the MapReduce Framework. A similar approach is taken by Yi Wang \etal in their work [28], where they focus specifically on processing-in-memory for CNNs. The primary objective of the greedy strategy proposed by us is comparable to some of the above-mentioned works, where the primary objective is to minimize the application execution times.

\subsubsection{Comparison of IPC \& Execution Time}
Fig. \ref{MAB_IPC} presents normalized IPC for all the four resource allocation strategies, i.e., random, grouped, greedy, and our MAB approach. Here the normalization is done with reference to the random resource allocation strategy. Only for two benchmarks - PARSEC-dedup and SPLASH2-raytrace, we observe grouped allocation strategy performing better than random allocation strategy by $1.275 \%$ and $3.5\%$ respectively. In all the other benchmarks, the performance of grouped allocation strategy is poorer by almost $2\%$ compared to the random allocation strategy. The performance of the grouped strategy plummets to nearly $15\%$ compared to the random strategy in the case of PARSEC-blackscholes. This can be justified mainly by the application features [43] [44], such as the granularity of instructions (fine, medium, or coarse), data usage, and use of the synchronization primitives (barriers, locks). However, the MAB strategy, across all the diverse applications studied, has performed better than all the resource allocation strategies by $\sim11\%$.  [43] [44]. We observe that specific benchmarks benefit much more from the MAB strategy like Black-Scholes, where the coarse granularity and low data usages help improve performance by nearly $19\%$.  

Occam's razor principle would suggest the greedy strategy to outperform the MAB resource allocation strategy in terms of performance metrics. However, we have tried to explain that in the case of real-world systems, it is not so. The best-performing cores are overused and stalled with long task queues, whereby some tasks which do not rely on the data can be put on execution on other underperforming cores and this task allocation can actually free up the system reducing the overall execution time. In the greedy strategy, the execution of the tasks was concentrated on the better-performing cores, which caused a significant stall period because of the task queue. The long-term reward approach in the MAB strategy balances the load across all the NMP cores, thereby performing better than the greedy.

\begin{figure}
    \centering
    \includegraphics[width=\textwidth]{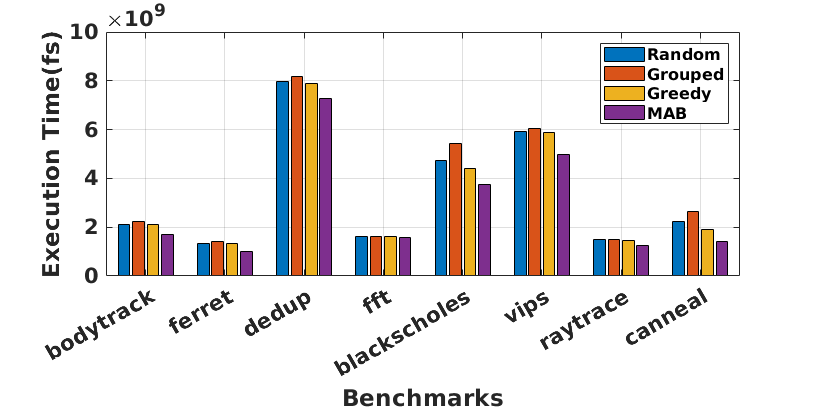}
    \caption{Performance Comparison in terms of Execution Time for Benchmarks from PARSEC and SPLASH2 Benchmarks Suite}
    \label{MAB_Execution_Time}
    \vspace{-0.5cm}
\end{figure}

Our work studied the execution times for the ROI, and these regions are the code sections where the tasks are offloaded to the NMP cores. Fig. \ref{MAB_Execution_Time} represents the execution time in femtoseconds. A reduced execution time highlights the improvement achieved by the MAB strategy with better mapping of tasks to the NMP cores. The study of IPC does not highlight the improvement brought out by the MAB strategy extensively in the case of the PARSEC-bodytrack and the PARSEC-vips, where the percentage improvement in IPC for the MAB strategy compared to the random strategy is $7.47\%$ and $5.5\%$ respectively. However, we can observe a considerable reduction in execution time for the ROIs such as $4.1 \times 10^8$fs, and $ 0.6 \times 10^8$fs reduction in execution time for PARSEC-bodytrack, and the PARSEC-vips respectively compared to random allocation. In terms of percentages, the MAB strategy reduces the execution time of the ROI by $19.4\%$ and $15.5\%$ respectively for PARSEC-bodytrack, and the PARSEC-vips, when compared to random allocation strategy. The MAB strategy could only achieve this optimization as it  includes the inter-task communication and the resource consumption cost (from equation \ref{eft_cost} and equation \ref{resource_cost}). 

\subsubsection{Study of L1 Cache at NMP cores}
\begin{figure} [t]
    \centering
    \includegraphics[width=\textwidth]{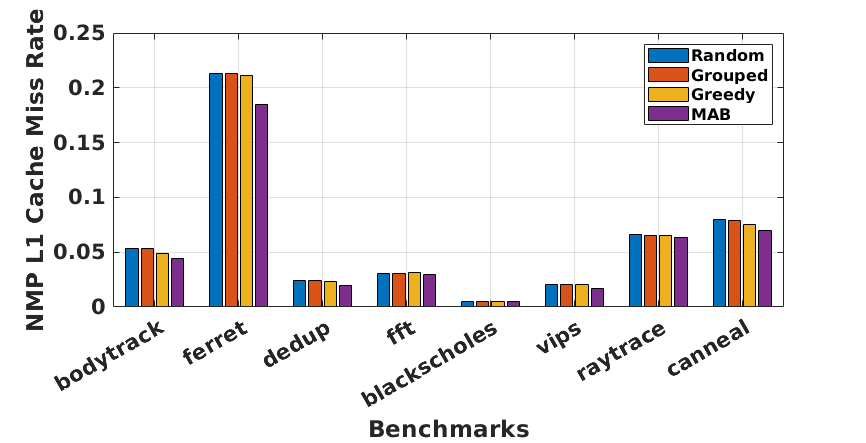}
    \caption{ Comparison of L1 Cache Miss Rate for PARSEC and SPLASH2 Benchmarks Suite}
    \label{MAB_miss_rate}
\end{figure}

Each NMP core is equipped with caches as presented in the NMP configuration table \ref{NMPSpec}. In fig. \ref{MAB_miss_rate} we offer the data for the cumulative L1 miss rates for all the NMP cores. As already discussed in section \ref{l1cacheoffload}, it is advisable to offload non-cacheable tasks (or tasks that thrive less on the locality of data) onto the NMP cores for execution. 

Intending to minimize the execution time on any of the NMP cores and reduce the inter-NMP core communication time, the MAB strategy itself learns which are the best tasks to be offloaded on to which of the NMP cores. Over the iterations, the learning strategy (Algorithm \ref{PIMCoreSelect}) can identify the NMP cores that serve the best for a particular task and then allocate the tasks to them. The immediate significant observation, which comes from fig \ref{MAB_miss_rate} is the reduced miss rates in the MAB strategy compared to all the other strategies under study. We observe very low L1 cache miss rates (values in the order of $10^{-2}$) are consistent for all the benchmarks under all the strategies. A certain number of mandatory misses which exist, could justify the reason for these low values. The marginal improvement in miss rates observed in the MAB strategy is happening only because of the better task allocation. In terms of percentages, PARSEC benchmarks ferret, bodytrack, and vips benefit by nearly $13\%, 16.8\%$, and $18.4\%$ respectively, compared to the random allocation strategy. If cache misses had increased, our MAB strategy would have failed to meet the expected improvements altogether.

\subsubsection{Study of the Packet Latencies}
\begin{figure}
    \centering
    \includegraphics[width=\textwidth]{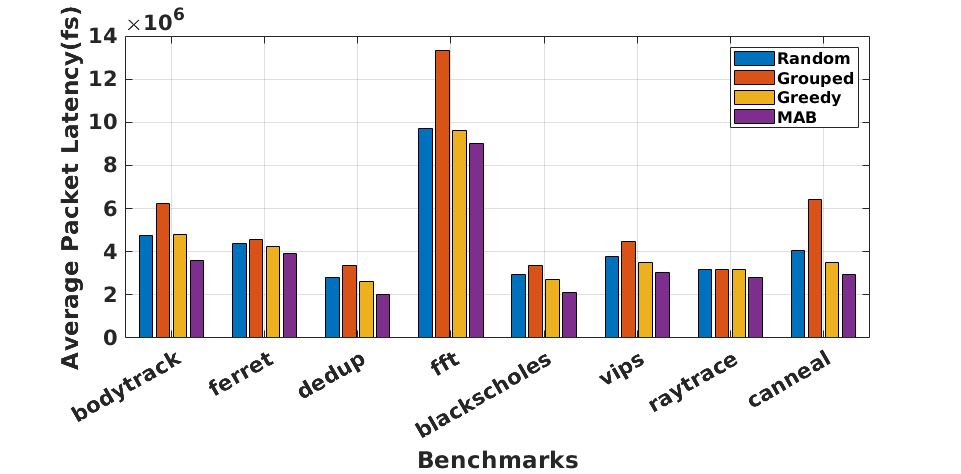}
    \caption{Comparison of average packet latency for Benchmarks from PARSEC and SPLASH2 Benchmarks Suite}
    \label{MAB_packet_lat}
    \vspace{-0.5cm}
\end{figure}

We measure the average traversal latency of the packets generated by the NMP cores under each 3D stacked memory unit (refer to section \ref{sim_env}). A resource allocation strategy is considered good if it creates lower contention and lower average packet latency. Fig. \ref{MAB_packet_lat} depicts the average packet latency for each of the benchmarks. Packets under the grouped resource allocation strategy tend to face higher contention because it prefers allocating tasks more often to a certain set of NMP cores in the MCN architecture and use the remaining cores based on the further requirement. The average packet latency in grouped resource allocation strategy for PARSEC-bodytrack, PARSEC-dedup, SPLASH2-fft, PARSEC-blackscholes, PARSEC-vips, and PARSEC-canneal is $1.471 \times 10^6$fs, $0.571 \times 10^6$fs,  $3.6 \times 10^6$fs, $0.449 \times 10^6$fs, $0.717 \times 10^6$fs, and $2.375 \times 10^6$fs respectively more than the random strategy. 

Observing the average packet latency after the implementation of the MAB strategy reduces the latency by $1.1 \times 10^6$fs, $0.76 \times 10^6$fs, $0.68 \times 10^6$fs, $0.813 \times 10^6$fs, $0.747 \times 10^6$fs, and $1.12 \times 10^6$fs for PARSEC-bodytrack, PARSEC-dedup, SPLASH2-fft, PARSEC-blackscholes, PARSEC-vips, and PARSEC-canneal respectively when compared to the random allocation strategy. Percentage improvement in the average packet latency because of the MAB strategy is between $5\%$ to $25\%$ across different benchmarks.
 
\subsubsection{Evaluating Total Power Consumption}
Table \ref{tab:power_conumed} presents the total power consumed by benchmarks under different resource allocation strategies and the maximum power reduced when the MAB strategy is applied. We have enabled Dynamic Voltage, and Frequency Scaling (DVFS) [5] for power saving when a core is not allocated any task. From equation \ref{reward}, we identify that the MAB strategy also considers the power density of each NMP core while allocating a task to it for execution. When a task is not allocated to any NMP, its power densities are regulated. Hence, the power cost is reduced, which helps the MAB strategy further decide over allocating future tasks to the specific NMP core. Thus, on average, the overall system power consumption improves by nearly $12\%$, and a maximum improvement of around $19.2\%$ is observed for the raytrace benchmark.

\begin{table}[!htbp]
\centering
\caption{Total power consumed in Watts by each benchmark under different allocation policy. Percentage improvement in power consumption for MAB strategy compared to random allocation strategy.}
\resizebox{\textwidth}{!}{{%
\begin{tabular}{|c|c|c|c|c|}
\hline
Benchmarks   & Random & Grouped & MAB    & \% Reduced (Max) \\ \hline
bodytrack    & 167.81 & 172.47  & 153.76 & 10.84            \\ \hline
ferret       & 164.64 & 166.10  & 151.76 & 8.6              \\ \hline
dedup        & 164.35 & 172.74  & 158.71 & 8.12             \\ \hline
fft          & 188.51 & 201.06  & 177.54 & 11.7             \\ \hline
blackscholes & 196.21 & 188.53  & 170.63 & 13.03             \\ \hline
vips         & 186.30 & 186.39  & 162.56 & 12.8             \\ \hline
raytrace     & 172.42 & 204.21  & 165.02 & 19.19            \\ \hline
canneal      & 176.33 & 170.62  & 161.45 & 8.44             \\ \hline
\end{tabular}%
}}
\label{tab:power_conumed}
\end{table}
\subsection{Overhead in the Benchmark Execution Time due to the MAB Algorithm}

 As seen in table \ref{time_compare}, the contribution of the MAB Algorithm to the execution time of the total program is nominal, almost of the order of $10^{-2}$. We observed that there is a fixed cost in terms of time consumed in the case of the MAB-based resource allocation strategy. This time  consumed by the MAB-Algorithm is to compute the reward and make the decision on which core is best suited for any particular task. The MAB strategy is usually limited by the re-iteration-based reward calculation.  
 
\begin{table}[!htbp]
\centering
\caption{MAB Algorithm execution time contribution to the benchmark execution time}
\resizebox{\textwidth}{!}{{%
\begin{tabular}{|c|c|c|l}
\cline{1-3}
Benchmark    & \begin{tabular}[c]{@{}c@{}}Total Execution Time per task\\ ($\times 10^{9}$ fs)\end{tabular} & \begin{tabular}[c]{@{}c@{}}MAB Algo. contribution \\ to Execution Time of per task\\ ($\times 10^{7}$ fs)\end{tabular} &  \\ \cline{1-3}
bodytrack    & 1.85 & 7.2                                                                                                                                    &  \\ \cline{1-3}
ferret       & 0.85                                                                                                            & 6.92                                                                                                                                   &  \\ \cline{1-3}
dedup        & 7.3                                                                                                             & 12.71                                                                                                                                  &  \\ \cline{1-3}
fft          & 1.66                                                                                                            & 7.17                                                                                                                                   &  \\ \cline{1-3}
blackscholes & 3.81                                                                                                            & 10.22                                                                                                                                  &  \\ \cline{1-3}
vips         & 5.03                                                                                                            & 11.37                                                                                                                                  &  \\ \cline{1-3}
raytrace     & 1.1                                                                                                             & 6.93                                                                                                                                   &  \\ \cline{1-3}
canneal      & 1.43                                                                                                            & 6.99                                                                                                                                   &  \\ \cline{1-3}

\end{tabular}
}}
\label{time_compare}

\end{table}

\subsection{Case Studies}

In this part of the section, we look into two of the benchmarks from two different domains.
A summary of the inherent characteristics of the Black-Scholes (Financial Analysis) and Canneal (Engineering) is presented in table \ref{parsec-bench-table}. This diversity in the inherent key characteristics of the two applications encourages us to pursue them for further evaluation.

\subsubsection{Black-Scholes Benchmark Case Study} \label{blackscholes_profile_case}

\begin{table*}[!htbp] \label{parsec-bench-table}
\centering
\caption{Inherent key characteristics of the Black-Scholes and Canneal [43]}
{%
\begin{tabular}{|c|cc|c|cc|ll}
\cline{1-6}
{Program} & \multicolumn{2}{c|}{Parallelization} & {\begin{tabular}[c]{@{}c@{}}Working \\ Set\end{tabular}} & \multicolumn{2}{c|}{Data Usage} &  &  \\ \cline{2-3} \cline{5-6}
 & \multicolumn{1}{c|}{Model} & Granularity &  & \multicolumn{1}{c|}{Sharing} & Exchange &  &  \\ \cline{1-6}
Black-Scholes (Finance) & \multicolumn{1}{c|}{data parallel} & coarse & small & \multicolumn{1}{c|}{low} & low &  &  \\ \cline{1-6}
Canneal (Engineering) & \multicolumn{1}{c|}{ustructured} & fine & unbounded & \multicolumn{1}{c|}{high} & high &  &  \\ \cline{1-6}
\end{tabular}%
}

\label{parsec-bench-table}
\end{table*}

\begin{figure}[t]
    \centering
    \includegraphics[width=\textwidth]{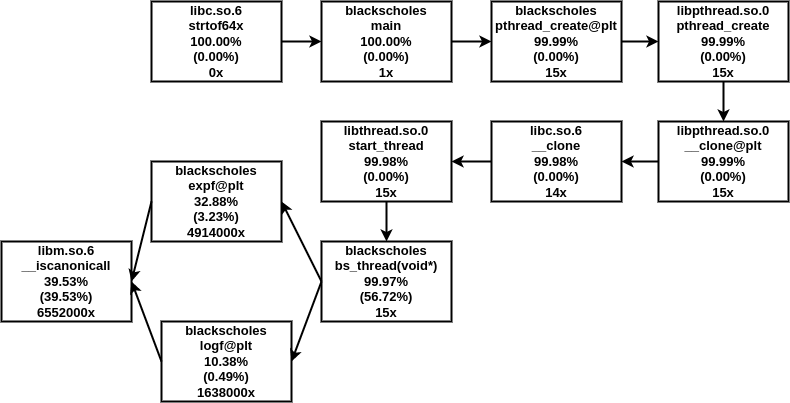}
    \caption{Profile of Black-Scholes Benchmark}
    \label{blackscholes_profile}
\end{figure}
In this section, we further look into a PDE from Financial Analysis domain. The Black-Scholes equation is a partial differential equation, which describes the price of the option over time. The equation is:
\begin{equation}
    \frac{\partial V}{\partial t}+\frac{1}{2}\sigma^2S^2\frac{\partial^2V}{\partial S^2}+rS\frac{\partial V}{\partial S}-rV=0 \nonumber
\end{equation}
where $V$ is an option on the underlying $S$ with volatility $\sigma$ at time $t$ if the constant interest rate is $r$. There exists no closed formed solution for the Black-Scholes equation and has to be solved numerically. 
\begin{figure*} [t]
    \centering  
\subfigure[Before Improvement]{\label{blackscholes_before}\includegraphics[width=0.45\textwidth]{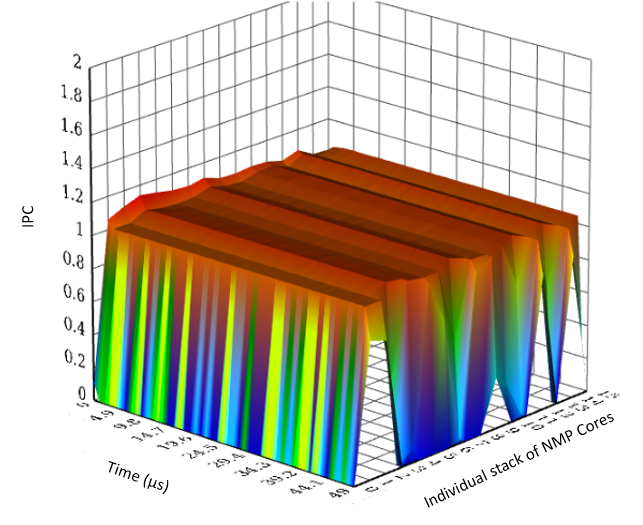}}
\subfigure[After Improvement]{\label{blackscholes_after}\includegraphics[width=0.45\textwidth]{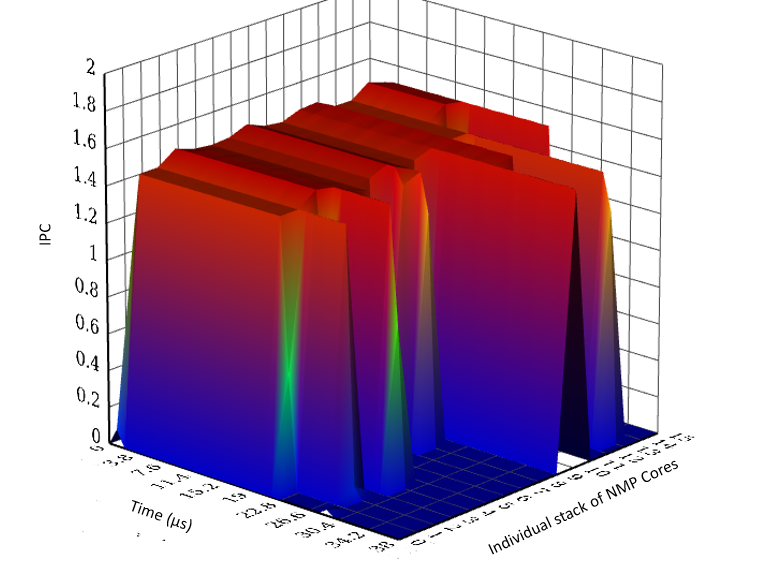}}
\caption{Black-Scholes Benchmark performance visualization over time for sections of code offloaded to group of NMP cores in each 3D stacked memory unit. The visualization clearly depicts the performance improvement over the time in terms of IPC and simultaneously depicts the cores in operations. Clearly instead of exhaustively using all the cores a better offloading reduces the execution time form $49\mu s$ to $38\mu s$ for the region of our interest.}
\label{blackscholes_perf}
\end{figure*}
The program is limited by the number of floating-point operations executed concurrently. Since, there are no data dependencies in the applications, the tasks can be fixed as predefined block size [34]. In fig \ref{blackscholes_perf}, we present a performance visualization for Black-Scholes in terms of IPC for a group of 16 cores under each 3D stacked memory unit. As shown in fig \ref{blackscholes_profile}, a portion of code is executed over a million times. Since the code section is not limited by any locality and majorly depends on performing a certain number of floating-point operations, we offload this code section to the NMP side for execution and observe the performance. This code section is also our ROI in the Black-Scholes benchmark as we implement our strategies. 

The program divides the work unit equally into the number of cores available and then executes them concurrently. Fig. \ref{blackscholes_before} depicts the performance for this region of interest and fig. \ref{blackscholes_after} represents the performance when the MAB strategy is applied to it. Better offloading and resource allocation increase the performance on the NMP, which can be observed from the IPC improvement from 1.1 to 1.5, with a simultaneous reduction in the time spent in executing the code section from 49$\mu s$ to 38$\mu s$. From table 4, we also observe that around $13\%$ improvement in power consumption is observed from pre-MAB to post-MAB strategy implementation. Fig. \ref{blackscholes_after} clearly shows how a group of NMP cores under 3D memory unit numbers 10, 11, and 13 are utilized. The objective of minimizing the regret by efficiently using the cores is achieved (cost is shown in equation 7 and regret optimization is shown in equation 9).  The MAB strategy potentially finds these cores capable of completing the computation with relatively minimal cost than selecting other cores or distributing the computation over all the other cores.


 \subsubsection{Canneal Benchmark Case Study} \label{canneal_profile_case}

\begin{figure}[t]
    \centering
    \includegraphics[width=0.7\textwidth]{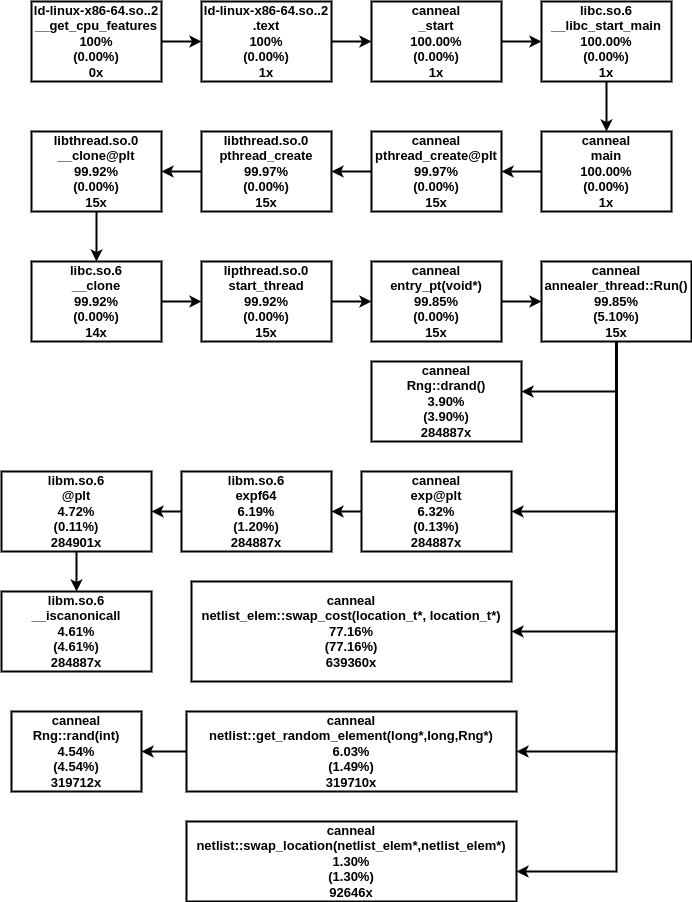}
    \caption{Profile of Canneal Benchmark}
    \label{canneal_prof}
\end{figure}

Canneal uses cache-aware simulated annealing (SA) to reduce the routing cost of System-on-Chip (SOC) design. The cache-aware simulated annealing randomly picks a pair of elements and swaps them. The algorithm discards only one element during each iteration to increase data reuse, effectively reducing capacity misses in caches. The SA method accepts swaps that increase the routing cost with a certain probability of making an escape from potential local minimal possible. This probability continuously decreases during runtime to allow the design to converge. The program represents engineering workloads, for the fine-grained parallelism with lock-free synchronization techniques and pseudo-random worst-case memory access pattern, making it diverse from the study of Black-Scholes. The selection of code sections in the case of fine-grained applications can be a real challenge. In canneal, the application randomly chooses a pair, deferences the pointers, and atomically swaps the two elements. The challenge in lock-free synchronization is that some other threads select any element on the pair within the dereferencing. With reference to the profile figure shown in fig. \ref{canneal_prof}, certain functions are offloaded to the NMP cores for execution, such as the $get\_random\_element, swap\_locations, swap\_cost$, and floating-point calculations; these code sections are also our ROI for the case study. 

\begin{figure*}[t]
    \centering 
\subfigure[Before Improvement]{\label{canneal_profile_before}\includegraphics[width=0.45\textwidth]{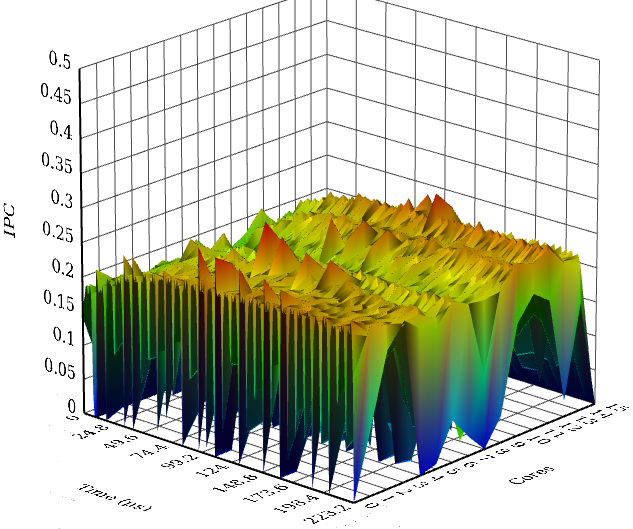}}
\subfigure[After Improvement]{\label{canneal_profile_after}\includegraphics[width=0.45\textwidth]{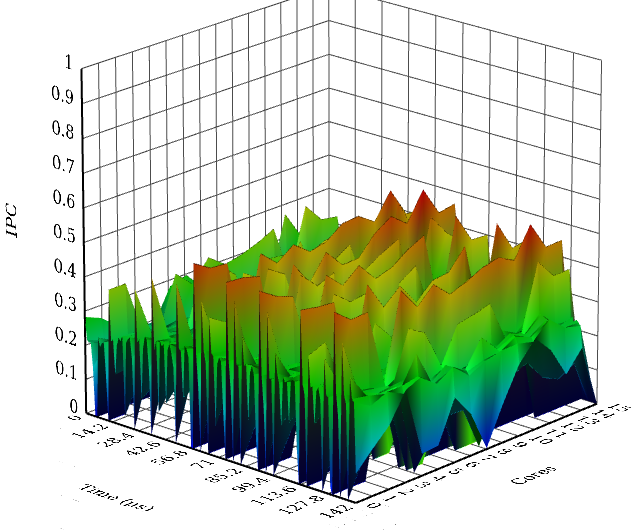}}

\caption{Canneal Benchmark performance visualization over time for sections of code offloaded to group of NMP cores in each 3D stacked memory unit. The visualization clearly depicts the performance improvement with reduced execution time for the region of interest. Canneal has higher inter-task communication and so the task offloading becomes more important. }
\label{canneal_perf}
\end{figure*}

Fig. \ref{canneal_perf} presents the performance visualization of each group of NMP cores under the 3D stacked memory structure. Fig. \ref{canneal_profile_before} and \ref{canneal_profile_after} are used to present the performance improvement of Canneal Benchmark before and after the application of the MAB approach. From table \ref{parsec-bench-table}, the working set of the Canneal benchmark is so large that it is considered unbounded. In fig. \ref{canneal_profile_before} all the available cores are very exhaustively used. Whereas in fig. \ref{canneal_profile_after} all the cores are active; however, some of the better performing cores with relatively lower costs (with reference to equation 4, and equation 6) are more extensively used. This can be visualized from the relatively higher IPC peaks achieved by some of the cores at a certain time. From table \ref{parsec-bench-table}, the Canneal Benchmark has higher data usage and have fine granularity (Granularity $=\frac{T_{comp}}{T_{comm}}$, where $T_{comp}$ is the computational time and $T_{comm}$ is the communication overhead time), causing the high inter-task communication. The regret optimization in the MAB strategy also considers the communication time and the task execution time. It helps improve the performance by improving the execution time for the region of interest by nearly 1.6 times. Also, the power consumed in applying the MAB strategy is 8.4\% less than the random allocation strategy.

\section{Conclusion} \label{conclusion}
In this paper we propose a multi-armed bandit approach using regret optimization to address the resource allocation problem. Our work is the first to study and propose the resource allocation strategy for an MCN architecture. We have studied multiple metrics such as IPC, execution time, cache miss rates at NMP core, average packet latency and percentage power reduction, to establish the overall improvement MAB strategy brings to the MCN architecture. The advantage in having Multi-Armed bandit Problem is that it highlights that the long term rewards benefit the system more than short term exploitative rewards which most of the current research work relies upon. We hope that our strategy motivates for further full reinforcement learning based resource allocation strategies in multi-stack NMP architectures.

 \appendix
 \section{Example NMP Offloaded code sections} \label{appendixA}
We have created regions of interest which are the code sections that can be offloaded to the NMP side. The decision is made on the code sections, which do not thrive much on the data locality when compared to other portions and can be parallel executed, thereby reducing the time spent in execution. We have used a wide variety of benchmarks from different application domains. The division of tasks varies widely from benchmark to benchmark; for example in the case of the Black-Scholes Benchmark, used in Financial Analysis, there are hardly any data dependencies to be taken care of, and tasks are made of predefined blocks. Another example of a benchmark that we have used is the Dedup. The Dedup application compresses data locally and globally in order to achieve higher compression rate. The application stages are broadly divided into 5 stages - Fragment, Fragment refine, Deduplication, Compress and Reorder. The Fragment Refine stage divides the data into fine grained granularity and the partitioned data ranges from few hundred to few thousands. By grouping and pipelining the stages correctly and ensuring the synchronization is done using locks, these tasks are executed on the NMP cores. In case of Canneal Benchmark, Canneal uses cache-aware simulated annealing (SA) to reduce the routing cost of System-on-Chip (SOC) design. The cache-aware simulated annealing randomly picks a pair of elements and
swaps them. The algorithm discards only one element during each iteration to increase data reuse, effectively reducing
capacity misses in caches. The SA method accepts swaps that increase the routing cost with a certain probability of
making an escape from potential local minimal possible. This probability continuously decreases during runtime to
allow the design to converge. The program represents engineering workloads, for the fine-grained parallelism with
lock-free synchronization techniques and pseudo-random worst-case memory access pattern, making it diverse from
the study of Black-Scholes. The selection of code sections in the case of fine-grained applications can be a real
challenge. In canneal, the application randomly chooses a pair, deferences the pointers, and atomically swaps the two
elements. The challenge in lock-free synchronization is that some other threads select any element on the pair within
the dereferencing. With reference to the profile figure shown in fig. \ref{canneal_prof}, certain functions are offloaded to the NMP
cores for execution, such as the get random element, swap locations, swap cost, and floating-point calculations;
these code sections are also our ROI for the case study. 

\begin{figure}[!htbp]
    \centering
    \includegraphics[width=0.5\textwidth]{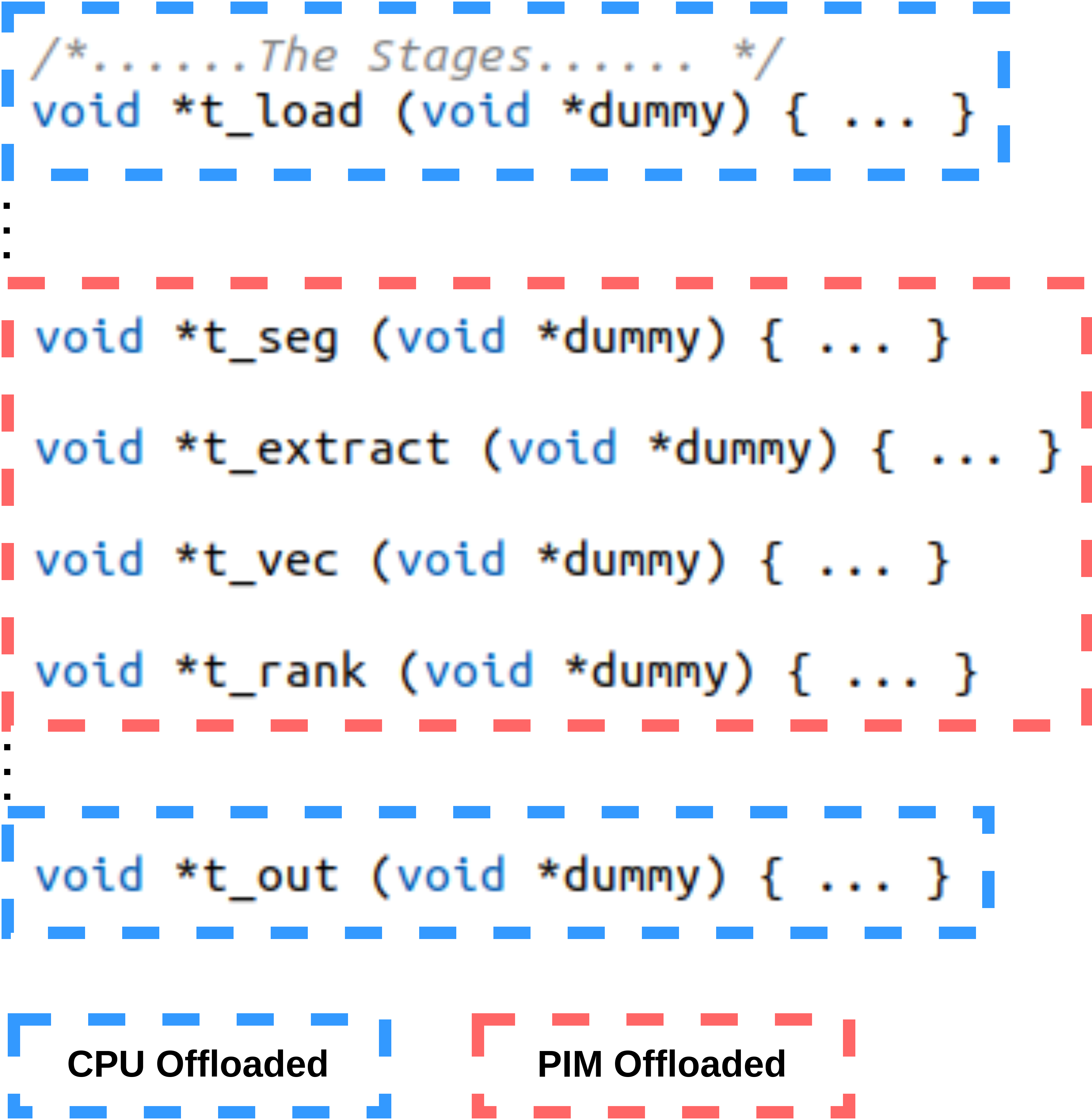}
    \caption{Code Profile of Ferret Benchmark for Offloading}
    \label{ferret_profile}
\end{figure} 

Ferret benchmark is configured for the image similarity search. The serial query can be broken into pipeline stages namely, load, segmentation, extract, vectorization, Rank and Output. Once the load stage, loads the image for query, the pipeline has only five stages, and there is no data dependencies between the individual queries. However, the $t\_out$ shares a common output file. Therefore, we offload the middle four stages to the NMP side for reliable execution.

\includepdf[pages=-]{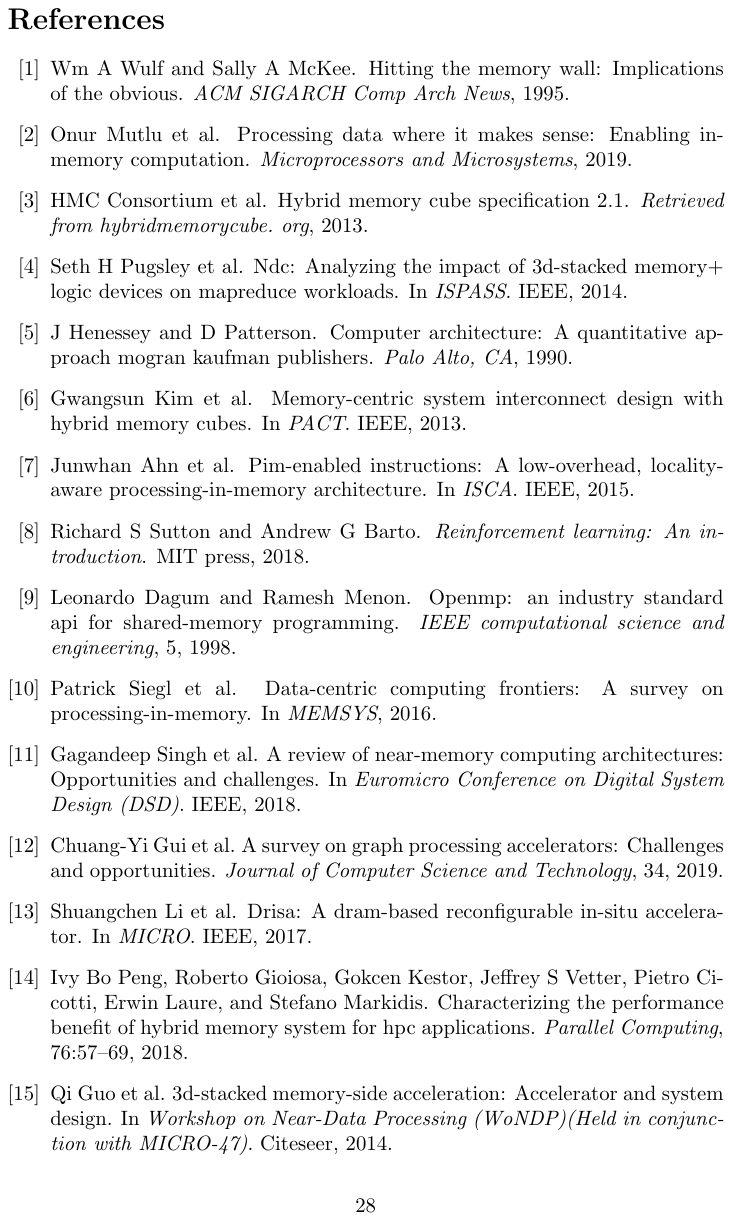}
\end{document}